# On the optimization of the magneto-plasma compressor power supply system

K.I. Deshko[a], V.A. Chernikov[b]

*M.V.Lomonosov Moscow State University. Moscow 119991, Russia.*

E-mail:[a] kir.deshko@gmail.com     [b] vachernikov@rambler.ru

The work of the miniature magneto-plasma compressor with the low-voltage power supply system was studied experimentally and the current and discharge voltage drop were measured. It was found that the voltage drop (a few tens of volts) remains practically constant during discharge. Performed electrical analysis allowed to determine the power efficiency of the supply system depending on the initial capacitor voltage, the discharge voltage and circuit parameters (capacitance, inductance, resistance).

# Об оптимизации системы питания магнитоплазменного компрессора

К.И. Дешко[a], В.А. Черников[b]

*Московский государственный университет имени М.В.Ломоносова,*

*физический факультет, кафедра физической электроники.*

*Россия, 119991, Москва, Ленинские горы, д. 1, стр. 2.*

E-mail:[a] kir.deshko@gmail.com    [b] vachernikov@rambler.ru

Экспериментально исследована работа миниатюрного магнитоплазменного компрессора с низковольтной системой питания; измерены ток и падение напряжения на разряде. Установлено, что во время разряда падение напряжения остается практически постоянным и составляет порядка нескольких десятков вольт. Проведенный электротехнический анализ позволил определить КПД системы питания в зависимости от начального напряжения накопителя, напряжения горения разряда и параметров разрядной цепи (ёмкости, индуктивности, сопротивления).

*Ключевые слова:* МПК, магнитоплазменный компрессор, импульсный плазматрон, оптимизация системы питания

УДК: 533.9.07

## Введение

Простота конструкции магнитоплазменного компрессора (МПК) и возможность получения высокоскоростных струй сильноионизованной плазмы [1] вызвали значительный интерес и привели к большому числу экспериментальных и теоретических работ, посвященных исследованию возможности применения МПК в различных сферах (как источник плазмы, оптического излучения, ударных волн, нейтронов; протекание ядерных реакций в плазменном фокусе) и определению параметров создаваемой им плазменной струи (пространственные и временные распределения концентраций и температур компонентов плазмы). Было продемонстрировано успешное применение МПК в качестве высокояркостного оптического источника в видимом и УФ диапазонах [2-8], инжектора для плазменных размыкающих ключей [9-13], формирователя ударно-сжатого слоя плазмы для обработки поверхностей материалов [14-16], запального устройства для поджига газотопливных потоков [17,18], источника нейтронов и рентгеновского излучения [19-21] и др.

При этом для дальнейшего применения МПК в науке и технике неизбежно встает задача оптимизации конструкции плазматрона и системы его питания с целью уменьшения массы, габаритов и повышения КПД.

Практически во всех экспериментальных работах питание МПК осуществлялось от высоковольтных конденсаторных батарей (емкость десятки – сотни микрофарад, начальное напряжение – единицы – сотни киловольт), что позволяло обеспечивать требуемый энерговклад и амплитуду тока в разряде.

Однако использование высоковольтных емкостных накопителей и повышающих зарядных устройств приводит, как минимум, к значительным массе и габаритам системы питания плазматрона. Это допустимо для лабораторных экспериментов, однако во многих применениях является неприемлемым и ограничивает использование МПК на практике.

Одним из методов решения указанных проблем является переход к низковольтной системе питания МПК. При этом уменьшение массы и габаритов достигается за счет отказа от повышающих зарядных устройств и использования низковольтных конденсаторов, обладающих большей удельной емкостью.

Настоящая работа посвящена исследованию работы МПК с низковольтной системой питания. На основе уравнений электротехники был выполнен анализ, позволивший определить энерговклад и КПД системы питания в зависимости от начального напряжения накопителя, напряжения горения разряда и параметров разрядной цепи (емкости, индуктивности, сопротивления).

## 1 Описание экспериментальной установки

Для экспериментального определения величины напряжения, падающего на МПК во время горения разряда, а также зависимости этого напряжения от протекающего в разрядной цепи тока и от параметров рабочей среды плазматрона (т.е. от давления и состава газа) была разработана и собрана экспериментальная установка, структурная схема которой представлена на рис. 1.

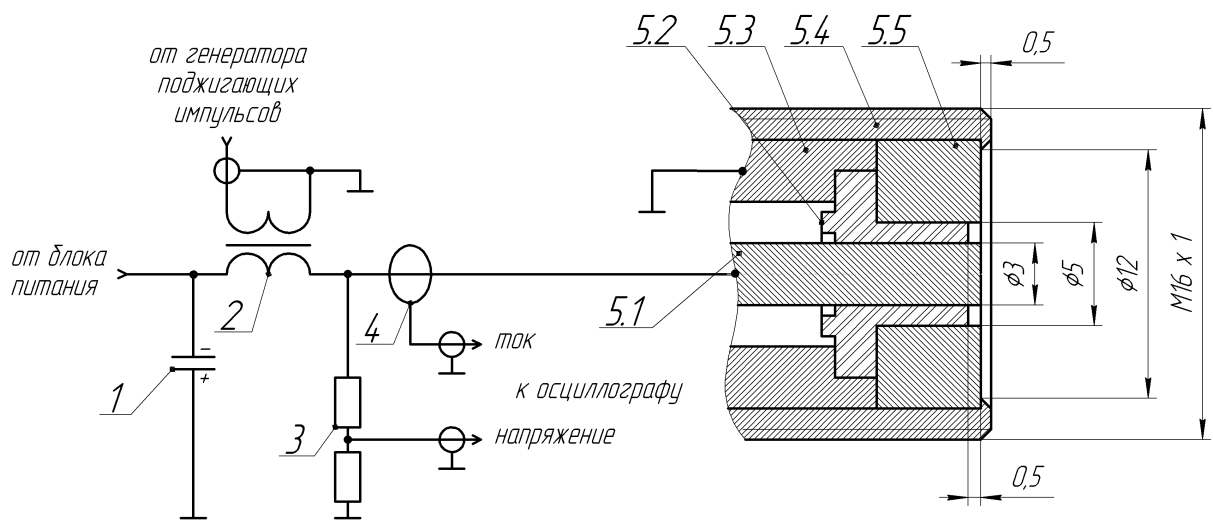

**Рис. 1.** Конструкция МПК и системы его питания.

1 – накопительный конденсатор, 2 – трансформатор последовательного поджига, 3 – делитель напряжения, 4 – пояс Роговского, 5 – МПК: 5.1 – центральный электрод (штырь из лантанированного вольфрама), 5.2 – диэлектрический разделитель из оксида циркония, 5.3 – внешний проводник подводящего коаксиала, 5.4 – корпус, 5.5 – периферийный электрод (шайба из лантанированного вольфрама).
На данном рисунке рабочий торец плазматрона – правый.

Установка состоит из накопительного конденсатора (1), заряжаемого от блока питания, трансформатора последовательного поджига (2), делителя напряжения (3) и пояса Роговского (4) для регистрации напряжения и тока. Магнитоплазменный компрессор (МПК) (5) был размещен в герметичной камере, подключенной к системе откачки и напуска газа, которая служила для предварительной откачки системы (минимальное давление не выше $1\times10^{-5}$ Торр), а также дозированного напуска рабочего газа (воздуха). Хотя исследования проводились в достаточно широком диапазоне давлений ($5\times10^{-1}...5\times10^{2}$ Торр), большинство измерений проведено при давлении в камере 50 Торр.

Блок питания обеспечивает зарядку накопительной емкости до заданного напряжения, величину которого можно было регулировать в пределах 0-300 Вольт. Накопительная емкость представляет собой батарею конденсаторов, для создания которой были использованы специальные алюмооксидные («электролитические») конденсаторы, обладающие малыми эквивалентными последовательными индуктивностью и сопротивлением и способные отдавать большие импульсные токи без разрушения. Общая емкость батареи составила ~2500 мкФ.

Для управляемой инициации разряда в плазматроне использовалась система последовательного поджига. В цепь питания МПК последовательно включена вторичная обмотка трансформатора; на первичную же обмотку подается запускающий импульс от соответствующего генератора. Высоковольтный импульс, наведенный во вторичной обмотке, вызывает пробой плазматрона и, затем, разряд на него основного накопительного конденсатора. Характерные параметры наведенного импульса напряжения: амплитуда ~10 кВ, длительность ~10 нс. Этих величин оказывается вполне достаточно, чтобы обеспечить устойчивый пробой плазматрона в воздухе при давлении в пределах от $10^{-1}$ Торр до атмосферного.

Эскиз конструкции МПК показан на рис. 1, там же отмечены основные размеры, определяющие геометрию плазматрона. Оба электрода – центральный, в виде штыря, и периферийный, в виде шайбы – были выполнены из лантанированного (2%) вольфрама. Выбор материала определялся соображениями эрозионной стойкости. В качестве диэлектрика, разделяющего электроды, использовалась керамическая трубка из оксида циркония. Опыт показал вполне удовлетворительную стойкость всех элементов плазматрона.

Величина тока в цепи питания плазматрона определялась при помощи калиброванного пояса Роговского, работавшего в режиме трансформатора тока и также встроенного в коаксиальный тракт.

Для измерения падения напряжения на МПК использовался делитель напряжения, подключенный максимально близко к плазматрону. Время нарастания переходной характеристики делителя составляло менее 50 нс, а постоянная времени разряда накопителя через делитель превышала 1 минуту.

Сигналы с пояса Роговского и делителя напряжения подавались на входы цифрового запоминающего осциллографа.

**2 Экспериментальные результаты**

Осциллограммы тока и падения напряжения на разряде МПК, полученные при различных начальных напряжениях на накопительном конденсаторе, показаны на рис. 2. Все они получены при разряде МПК в воздухе при давлении в вакуумной камере 50 Торр.

Из представленных осциллограмм видно, что после зажигания разряда ток довольно быстро (~10 мкс) нарастает до своего максимального значения порядка нескольких килоампер, а затем следует медленный (постоянная времени порядка 100 мкс) его спад. Видно, что с ростом начального напряжения временной характер изменения разрядного тока практически не изменяется. Растет только величина максимального значения тока.

В то же время из осциллограммы напряжения следует, что до зажигания разряда сопротивление межэлектродного зазора МПК велико, и все напряжение накопителя падает на нем. После пробоя и зажигания разряда напряжение достаточно быстро (примерно за то же время, в течение которого ток нарастает до своего амплитудного значения, т.е. порядка 10 мкс) уменьшается до некоторого стационарного значения, не зависящего от протекающего тока. В описываемых экспериментах это значение составляет величину порядка 30 Вольт.

Экспериментально было установлено, что при изменении давления газа в рабочей камере плазматрона в широком диапазоне $(5\times10^{-1}...5\times10^{2})$ Торр падение напряжения на разряде остается постоянным (в пределах ±5%), кроме того, из полученных результатов следует, что изменение давления газа не сказывается на токе разряда.

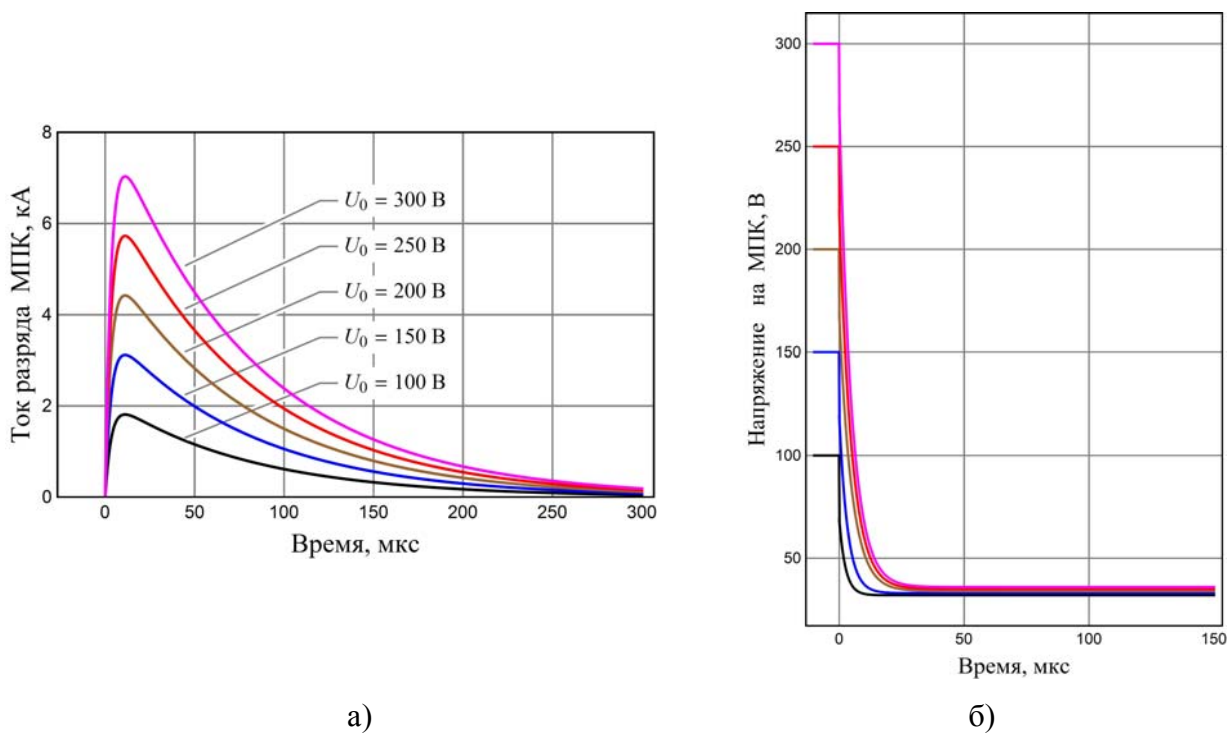

а)  б)

**Рис. 2.** Осциллограммы тока разряда (а) и падения напряжения на МПК (б) при различных начальных напряжениях накопителя.

### 3 Оптимизация системы питания на основе электротехнического анализа

Эквивалентная схема системы питания МПК может быть представлена в виде $rLC$ – цепи, как это показано на рис. 3,а. Проведем анализ процессов, протекающих в такой цепи, с точки зрения электротехники. В момент начала разряда ($t = 0$) накопительная емкость C замыкается на последовательно соединенные сопротивления $r_1$ (активное сопротивление разряда МПК), $r_2$ (совокупное омическое сопротивление разрядной цепи) и совокупную индуктивность $L$.

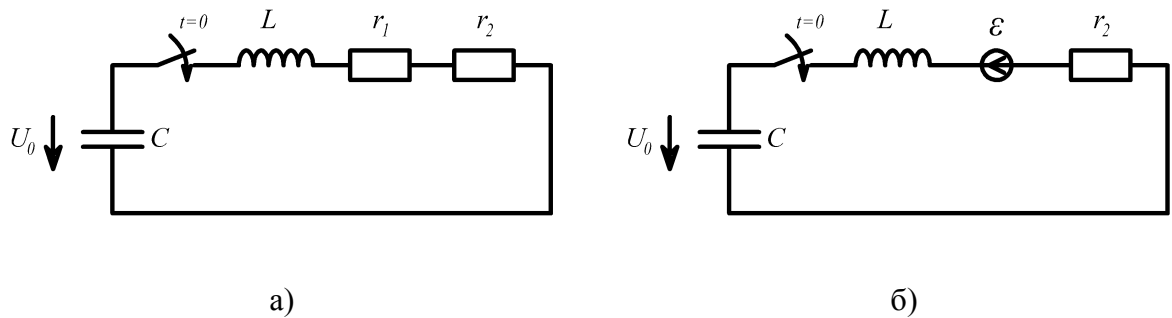

а)  б)

**Рис. 3.** Схема замещения системы питания МПК.

Будем считать, что практически сразу же после зажигания разряда падение напряжения на МПК достигает постоянного значения (согласно экспериментальным данным, рис 2,б). Исходя из этого упрощения, для удобства расчетов в соответствии с теоремой компенсации сопротивление МПК $r_1$ можно заменить на источник постоянной ЭДС ε (см. рис. 3,б).

Обычно при анализе переходных процессов в последовательной $rLC$ – цепи выделяют три характерных режима [19]:

1. *Апериодический режим*, реализующийся при $L < \frac{r^2 C}{4}$;

2. *Критический режим*, соответствующий переходу от апериодического к периодическому (колебательному) при $L = \frac{r^2 C}{4}$;

3. *Периодический режим* при $L > \frac{r^2 C}{4}$.

### 3.1 Апериодический режим

В соответствии со вторым законом Кирхгофа для тока в контуре на рис. 3,б имеем: $r_2 i(t) + L \frac{di(t)}{dt} + \frac{1}{C} \int i(t) dt - \varepsilon = 0$, откуда с учетом начальных условий и законов коммутации $i(t) = \frac{(U_0 - \varepsilon)}{L\sqrt{\delta^2 - \omega_0^2}} e^{-\delta t} \text{sh}(\sqrt{\delta^2 - \omega_0^2} t)$, где введены стандартные обозначения $\delta = \frac{r_2}{2L}$ и $\omega_0 = \frac{1}{\sqrt{LC}}$.

Условие погасания разряда можно определить из следующих соображений:
1. Выделение энергии на МПК всегда диссипативно, т.е. энергия, выделившаяся на МПК, не может вернуться в электрическую форму;
2. Известно [20], что для дугового разряда существует пороговый ток, при котором еще может гореть дуга; если же внешняя схема становится не в состоянии обеспечить такой ток, дуговой разряд гаснет. Величина порогового тока, по данным работы [20], для употребляемых на практике металлов составляет от 0.04 А (ртуть) до 6 А (никель). Для меди и вольфрама, наиболее часто используемых для изготовления электродной системы МПК, величина порогового тока не превышает 1.6 А Амплитуда тока в разряде МПК обычно составляет от единиц килоампер и выше, поэтому при анализе апериодического режима величиной порогового тока можно пренебречь и считать, что разряд длится бесконечно долго.

Энергия, запасенная в накопительной емкости к началу разряда $E_0 = \frac{CU_0^2}{2}$; после разряда останется $E_f = \frac{C\varepsilon^2}{2}$. За время разряда на МПК выделится энергия $E_1 = \varepsilon \int_0^\infty i(t)dt = C\varepsilon(U_0 - \varepsilon)$. Энерговклад в разряд, таким образом, пропорционален $(U_0 - \varepsilon)$.

КПД системы питания можно определить двумя разными способами. При однократном разряде он определяется как отношение выделившейся на МПК энергии к изначально запасенной и составляет $\eta_1 = \frac{E_1}{E_0} = \frac{2\varepsilon(U_0 - \varepsilon)}{U_0^2}$. При повторяющемся разряде, с учетом энергии, остающейся в накопителе $\eta_2 = \frac{E_1}{E_0 - E_f} = \frac{2\varepsilon}{U_0 + \varepsilon}$. Отметим, что в оба выражения входят только начальное напряжение накопителя и падение напряжения на разряде МПК, т.е. КПД не зависит, например, от омических потерь в схеме, емкости накопителя или величины индуктивности.

Зависимости КПД системы питания в разовом и частотном режимах от начального напряжения на накопителе приведены на рис. 4. Отметим, что обе зависимости имеют физический смысл только при $U_0 > \varepsilon$, т.е. когда возможно зажигание разряда.

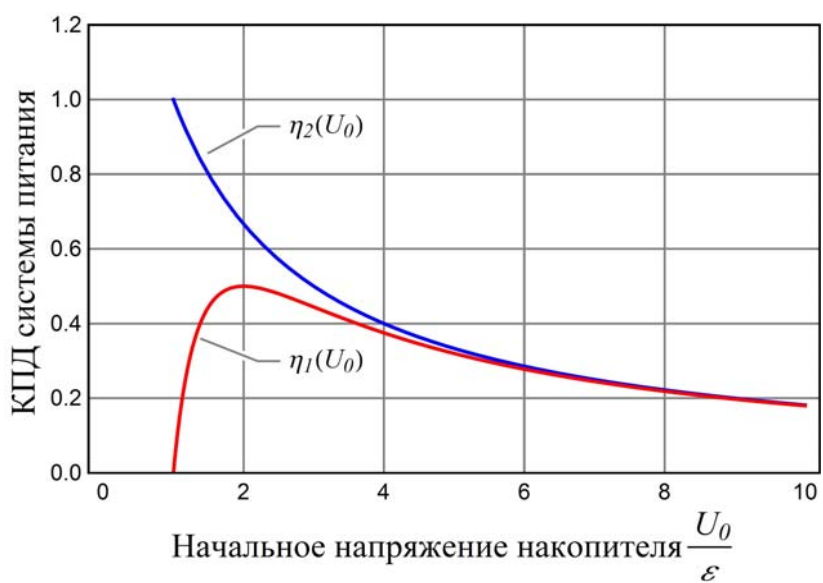

**Рис. 4.** Зависимости КПД системы питания в разовом ($\eta_1$) и частотном ($\eta_2$) режимах от начального напряжения накопительного конденсатора.

Начальное напряжение накопительного конденсатора, при котором достигается максимальный КПД при однократном разряде, можно определить из условия $\dfrac{d\eta_1(U_0)}{dU_0} = 0$, откуда оптимальное значение $U_0 = 2\varepsilon$. Достигаемый при этом КПД составляет $\eta_{1\max} = 50\%$. Из этого результата следует, что _при апериодическом разряде невозможно передать на МПК энергию больше половины запасенной в накопителе._ При этом четверть изначально запасенной энергии остается в накопителе, а еще четверть – рассеивается на активном сопротивлении цепи.

### 3.2 Критический режим

Критический режим также весьма прост для анализа, поскольку в нем величины $r, L, C$ связаны соотношением $L = \dfrac{r^2 C}{4}$. Зависимость тока в контуре от времени находится так же, как и при апериодическом режиме – решением уравнения, полученного на основании второго закона Кирхгофа:

$$i(t) = \frac{(U_0 - \varepsilon)}{L} t e^{-\frac{r_2 t}{2L}} = \frac{4(U_0 - \varepsilon)}{r_2^2 C} t e^{\frac{-2t}{r_2 C}}.$$

Выделившаяся на МПК энергия

$$E_1 = \varepsilon \int_0^\infty i(t) dt = C\varepsilon(U_0 - \varepsilon),$$ т.е. точно такая же, как и при апериодическом режиме.

### 3.3 Периодический (колебательный) режим

При колебательном режиме ток в контуре меняет направление, и поэтому разряд не может длиться бесконечно долго. Пренебрегая током обрыва, можно считать, что разряд погаснет при переходе тока через ноль. Вся оставшаяся электрическая энергия при этом будет запасена в емкости. В силу равенства тока нулю в индуктивности энергии запасено не будет, и, следовательно, не будет и ЭДС самоиндукции. Если после погасания разряда напряжение на емкости будет больше $\varepsilon$, то разряд зажжется повторно, только при этом и направление тока, и направление ЭДС $\varepsilon$ будут противоположны тем, что были до обрыва. Таким образом, для анализа достаточно рассмотреть процесс разряда от момента зажигания до первого перехода тока через нуль, поскольку последующие процессы будут качественно аналогичны.

При периодическом режиме зависимость тока от времени

$$i(t) = \frac{(U_0 - \varepsilon)}{L\sqrt{\omega_0^2 - \delta^2}} e^{-\delta t} \sin(\sqrt{\omega_0^2 - \delta^2} \, t).$$ Момент первого погасания разряда $\tau = \frac{\pi}{\sqrt{\omega_0^2 - \delta^2}}$.

На МПК выделится энергия

$$E_1 = \varepsilon \int_0^\tau i(t) dt = C\varepsilon(U_0 - \varepsilon)\left(1 + e^{\frac{-\pi\delta}{\sqrt{\omega_0^2 - \delta^2}}}\right) = C\varepsilon(U_0 - \varepsilon)\left(1 + e^{\frac{-\pi}{\sqrt{\frac{4L}{r_2^2 C} - 1}}}\right).$$

В этом выражении множитель $C\varepsilon(U_0 - \varepsilon)$ (такой же, как и при асимптотическом или критическом режимах) описывает зависимость выделяемой на МПК энергии от начального напряжения, а множитель $\left(1 + e^{\frac{-\pi}{\sqrt{\frac{4L}{r_2^2 C} - 1}}}\right)$, который может принимать значения от 1 до 2, - зависимость от соотношения $r_2$, $L$ и $C$. Поэтому для анализа зависимости выделившейся энергии от $r_2, L, C, U_0$ можно рассмотреть два случая.

1. Параметры цепи (соотношение $r_2, L, C$) фиксированы, а начальное напряжение $U_0$ накопителя меняется. В этом случае качественно зависимость та же, что и для апериодического или критического случаев, однако количественно выделяемая на МПК энергия и, следовательно, КПД выше, поскольку при периодическом режиме $L > \dfrac{r_2^2 C}{4}$ и $\left(1 + e^{\frac{-\pi}{\sqrt{\frac{4L}{r_2^2 C} - 1}}}\right) > 1$.

2. Начальное напряжение накопителя фиксировано, варьируется соотношение $r_2, L, C$. Видно, что при $L \to \infty$ множитель $\left(1 + e^{\frac{-\pi}{\sqrt{\frac{4L}{r_2^2 C} - 1}}}\right)$ стремится к своему максимальному значению, равному 2. Это соответствует полному разряду накопителя. Амплитуда тока при этом стремится к нулю, а длительность разряда - к бесконечности. В этом проявляется действие индуктивности – она обеспечивает более полную передачу энергии от накопителя в МПК, но одновременно снижается амплитуда тока.

### 4 Сравнение экспериментальных и теоретических результатов

В рассмотренных выше экспериментах был реализован апериодический режим разряда: это видно из приведенных осциллограмм тока и напряжения (см. рис. 2). По осциллограммам тока были также определены величины эквивалентных индуктивности и омического сопротивления, составившие $L \approx 110$ нГн и $r_2 \approx 34$ мОм соответственно. При этом $\delta \approx 2.6\omega_0$. На основании полученных значений параметров системы питания МПК были выполнены расчеты временной зависимости тока разряда и сравнение полученных кривых с экспериментальными осциллограммами. Оказалось, что во всех случаях наблюдается вполне удовлетворительное совпадение осциллограмм тока с теоретическими расчетами.

Ранее было показано, что при неизменных $r_2, L, C$ амплитуда тока разряда пропорциональна разности начального напряжения накопителя и напряжения горения разряда: $I \propto U_0 - \varepsilon$. Это позволило определить напряжение горения разряда также из осциллограмм тока. Для этого была построена зависимость амплитуды тока от начального напряжения накопителя (рис. 5). Эта зависимость оказалась линейной (на рисунке точки соответствуют результатам эксперимента, прямая – аппроксимация методом наименьших квадратов). Напряжение горения разряда определялось графически, как точка пересечения прямой с осью абсцисс.

Описанный способ «косвенного» определения напряжения горения разряда может быть полезен в тех случаях, когда использование делителя напряжения затруднено.

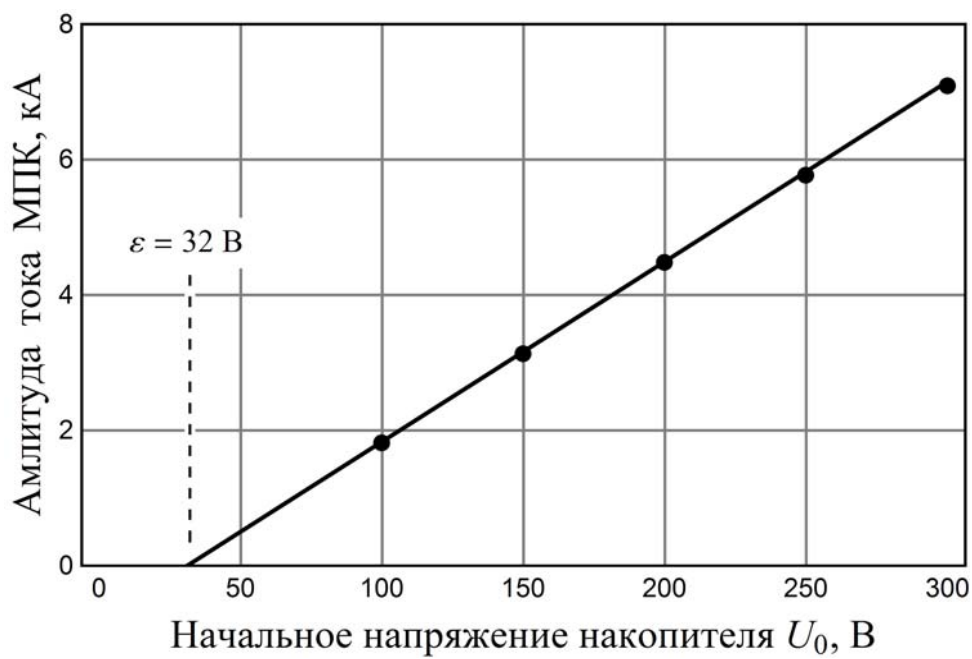

**Рис. 5.** Зависимость амплитуды тока МПК от начального напряжения накопителя. Точки – эксперимент, прямая – линейная экстраполяция.

**Выводы**

Экспериментально и теоретически исследовалась работа низковольтной системы питания МПК. На основании проведенных исследований были получены следующие основные результаты.

1. Низковольтная система питания МПК позволяет обеспечить работу плазматрона с такой же амплитудой разрядного тока (несколько килоампер), какая была получена при использовании высоковольтных систем [11,12,14,15].

2. После пробоя МПК и зажигания разряда падение напряжения на плазматроне составляет порядка нескольких десятков Вольт и не зависит от тока разряда;

3. При использовании воздуха в качестве рабочего газа падение напряжения также не зависит от начального давления в пределах $(5\times10^{-1}...5\times10^{2})$ Торр;

4. Полученные экспериментальные результаты позволили провести анализ работы системы питания с точки зрения электротехники и сделать следующие выводы:

- При любых величинах $C, L, r_2$ энерговклад (энергия, выделившаяся во время одного разряда на МПК) и амплитуда тока зависят от начального напряжения накопителя как $E_1, I \propto (U_0 - \varepsilon)$;

- При работе системы питания в апериодическом или критическом режиме энерговклад $E_1 = C\varepsilon(U_0 - \varepsilon)$ пропорционален емкости накопителя и не зависит от величин $L, r_2$, а КПД $\eta = \dfrac{2\varepsilon(U_0 - \varepsilon)}{U_0^2}$ не зависит от $C, L, r_2$. Максимальный КПД в разовом режиме составляет $\eta_{\max} = 50\%$ и достигается при $U_0 = 2\varepsilon$.

- При одних и тех же значениях накопительной емкости и начального напряжения накопителя энерговклад и КПД при работе схемы в периодическом режиме выше энерговклада и КПД в апериодическом или критическом режимах в $\left(1 + e^{\dfrac{-\pi}{\sqrt{\dfrac{4L}{r_2^2 C} - 1}}}\right)$ раз.

## Литература